# Did Saturn's Rings Form During The Late Heavy Bombardment?

**Sébastien CHARNOZ***
UMR AIM, Université Paris Diderot/CEA/CNRS
CEA/SAp, Centre de l'Orme Les Merisiers
91191 Gif-Sur-Yvette Cedex
France

**Alessandro MORBIDELLI**
Observatoire de la Côte d'Azur
OCA, B.P. 4229, 06304 Nice Cedex 4, France

**Luke DONES**
Southwest Research Institute
1050 Walnut St., Suite 300,
Boulder, Colorado 80302, USA

**Julien SALMON**
UMR AIM, Université Paris Diderot/CEA/CNRS
CEA/SAp, Centre de l'Orme Les Merisiers
91191 Gif-Sur-Yvette Cedex
France

Pages = 39

Figures = 9

Tables= 3

* : To whom correspondence should be addressed : charnoz@cea.fr

**Abstract :** The origin of Saturn's massive ring system is still unknown. Two popular scenarios - the tidal splitting of passing comets and the collisional destruction of a satellite - rely on a high cometary flux in the past. In the present paper we attempt to quantify the cometary flux during the Late Heavy Bombardment (LHB) to assess the likelihood of both scenarios. Our analysis relies on the so-called "Nice model" of the origin of the LHB (Tsiganis et al., 2005; Morbidelli et al., 2005; Gomes et al., 2005) and on the size distribution of the primordial trans-Neptunian planetesimals constrained in Charnoz & Morbidelli (2007). We find that the cometary flux on Saturn during the LHB was so high that both scenarios for the formation of Saturn rings are viable in principle. However, a more detailed study shows that the comet tidal disruption scenario implies that all four giant planets should have comparable ring systems whereas the destroyed satellite scenario would work only for Saturn, and perhaps Jupiter. This is because in Saturn's system, the synchronous orbit is interior to the Roche Limit, which is a necessary condition for maintaining a satellite in the Roche zone up to the time of the LHB. We also discuss the apparent elimination of silicates from the ring parent body implied by the purity of the ice in Saturn's rings. The LHB has also strong implications for the survival of the Saturnian satellites: all satellites smaller than Mimas would have been destroyed during the LHB, whereas Enceladus would have had from 40% to 70% chance of survival depending on the disruption model. In conclusion, these results suggest that the LHB is the "sweet moment" for the formation of a massive ring system around Saturn.

# 1. Introduction

The origin of Saturn's main ring system is still an unsolved question of modern planetary science. Whereas the origin of the rings of Jupiter, Uranus and Neptune, as well as of the dusty E and G rings of Saturn, seems to be linked to the presence of nearby moonlets (via their destruction or surface erosion, see Esposito 1993; Colwell 1994; Burns et al., 2001; Hedman et al., 2007; Porco et al., 2006), the unique characteristics of Saturn's main rings still challenge all scenarios for their origin. Saturn's main rings have a mass of the order of one to several Mimas masses (Esposito et al., 1983; 2007; Stewart et al., 2007) and are mainly composed of pure water ice, with only a few contaminants (Cuzzi & Estrada 1998; Poulet et al., 2003; Nicholson et al., 2005).

Historically, three main scenarios for the origin of Saturn's rings have been suggested and are still debated. They can be summarized as follows:

(1)     A satellite was originally in Saturn's Roche zone and was destroyed by a passing comet (Pollack et al., 1973, Pollack 1975, Harris 1984)
(2)     A massive comet (a Centaur object) was tidally disrupted during a close and slow encounter with Saturn (Dones 1991, Dones et al., 2007)
(3)     The rings are the remnants of Saturn's sub-nebula disk (Pollack et al., 1973, Pollack 1975).

The third scenario is less popular today, mainly because of the strong difference in the average chemical composition of Saturn's rings compared with Saturn's classical satellites (Harris 1984) which (in this scenario) should have originated from the same disk. In the present paper, we will deal only with scenarios #1 and #2, which strongly depend on the passage of one, or several, big "comets" very close to Saturn. A key question is: When might such events have been possible?

The rings' rapid evolutionary processes (viscous spreading of the A ring, Esposito 1986; surface darkening due to meteoritic impacts, Doyle et al., 1989; Cuzzi & Estrada 1998) have argued for a young ring system, perhaps less than $10^8$ years old. This contrasts with the fact that the current rate of passing comets is far too low for either of the scenarios above to have been likely during the last billion years of the Solar System history (Harris, 1984, Dones 1991, Lissauer et al. 1988). However, some recent re-evaluations and numerical modeling of ring evolutionary processes suggest that the rings may be older than previously thought: the viscosity in the dense rings seems now to be smaller than previously estimated, which results in longer spreading timescales (based on numerical modeling of gravity wakes, see Daisaka et al., 2001); Monte Carlo simulations of regolith growth show that an efficient re-surfacing of the ring particles is possible (Esposito et al., 2007), which may provide a solution to the surface darkening problem.

In addition, recent Cassini observations suggest the existence of macroscopic bodies in the ring system. More specifically, the detection of "propeller"-shaped structures in Saturn's A ring implies the presence of 50-100 m bodies in Saturn's A ring (Tiscareno et al. 2006, Sremcevic et al., 2007; Tiscareno et al. 2008) and the strange shapes of Pan and Atlas (Charnoz et al., 2007, Porco et al., 2007) show that 10-20 km bodies denser than ice (now covered with ring particles) were probably present during the formation of the rings. The existence of these macroscopic bodies gives new support to a scenario of ring origin through the catastrophic disruption of a massive progenitor.

So the debate over the age of Saturn's ring system is still open, and the possibility that the rings are as old as the Solar System must be considered seriously and examined in the light of recent advances in our understanding of Solar System formation and evolution.

Clearly, a key element for any formation scenario of Saturn's rings is the bombardment history of the giant planets. This has been investigated in several papers, with considerations based on the surface density and size distribution of the craters on the regular satellites of the giant planets (Smith et al., 1981, 1982; Lissauer et al., 1988; Zahnle et al., 2003). In Zahnle et al., (2003) the current impactor flux is derived from a model of the distribution of ecliptic comets (Levison & Duncan 1997, Levison et al. 2000). For the impactors' size distribution, functions compatible with either Ganymede's craters, Triton's craters or present day ecliptic comets are assumed. With these assumptions, Zahnle et al. derive a cratering rate on Iapetus that is between $2 \times 10^{-16}$ km$^{-2}$year$^{-1}$ and $8 \times 10^{-15}$ km$^{-2}$year$^{-1}$ (for craters with diameter D> 10 km). Integrated over the age of the Solar System, these rates would imply a current density of D>10 km craters that is between $10^{-6}$ km$^{-2}$ and $3.5 \times 10^{-5}$ km$^{-2}$. Smith et al., (1982) report a D>10 km crater surface density of about $2.3 \times 10^{-3}$ km$^{-2}$, and Neukum et al., (2005) report a surface density of about $8 \times 10^{-4}$ km$^{-2}$. These numbers are 100 to 2000 times larger than the value estimated in Zahnle et al., (2003), which argues that the present day comet-flux is much lower than in the past. In other words, it argues that the satellites of the giant planets, like our Moon, experienced an intense bombardment in the past. When this bombardment happened is difficult to say, in the absence of direct chronological measurements, However, the fact that the ejecta blankets of the basins on Iapetus overlap Iapetus' equatorial ridge, together with the model result that this ridge formed several hundred My after the accretion of the satellite (Castillo et al., 2007), suggest that the heavy bombardment of the giant planets satellites was late, as for the Moon.

Thus, in this paper we assume that the Late Heavy Bombardment (LHB) was a global event that concerned not only the Moon and the planets in the inner Solar System, but also the giant planets and their satellites, as predicted by the so-called "Nice model" (Tsiganis et al., 2005, Gomes et al., 2005, Morbidelli et al., 2005). In section 2 we briefly review the Nice model and in section 3 we use it to compute the collision probability of trans-Neptunian planetesimals with the objects in Saturn's system. In section 4 we discuss the possible size distributions of the impactors using constraints from (a) the Nice model, (b) the populations of comet- size objects in the Scattered Disk and in the Oort Cloud and (c) the crater record on Iapetus. With these premises, in section 5 we assess the likelihood that the comet tidal disruption scenario and the satellite collisional disruption scenario occurred during the LHB. In the last section, we discuss the pros and cons of each scenario, with considerations on the uniqueness of Saturn's ring system and on the missing silicate problem, i.e., the purity of the water ice in Saturn's rings.

## 2. The "Nice model" of the LHB

A comprehensive model for the origin of the LHB has been recently proposed. This model −often called the "Nice model" − quantitatively reproduces not only most of the LHB's characteristics (Gomes et al., 2005), but also the orbital architecture of the giant planet system: orbital separations, eccentricities, inclinations (Tsiganis et al., 2005), the capture of the Trojan populations of Jupiter (Morbidelli et al., 2005) and Neptune (Tsiganis et al., 2005; Sheppard and Trujillo 2006), the origin of the current structure of the Kuiper belt (Levison et al., 2008) and the capture of the irregular satellites of Saturn, Uranus and Neptune (Nesvorny et al., 2007). In the Nice model, the giant planets are assumed to be initially on nearly-circular and coplanar orbits, with orbital separations significantly smaller than those currently observed. More precisely, the giant planet system is assumed to lie in the region from ~5.5 AU to ~14 AU, and Saturn is assumed to be closer to Jupiter than their mutual 1:2 mean motion resonance. A planetesimal disk is assumed to exist beyond the orbits of the giant planets, on orbits whose dynamical lifetime is at least 3 My (the supposed lifetime of the gas disk). The outer edge of the planetesimal disk is assumed to be at ~34 AU and the disk's total mass is ~35 Earth masses. With the above configuration, the planetesimals at the inner edge of the disk evolve onto Neptune-scattering orbits on a timescale of a few million years. Consequently, the migration of the giant planets proceeds at a very slow rate, governed by the slow planetesimal escape rate from the disk. Because the planetary system would be stable in the absence of interactions with the planetesimals, this slow migration continues for a long time, slightly slowing down as the unstable disk particles are removed from the system. After a long time, ranging from 350 My to 1.1 Gy in the simulations of Gomes et al., (2005) – which is consistent with the timing of the LHB, approximately 700–750 My after planet formation – Jupiter and Saturn eventually cross their mutual 1:2 mean-motion resonance. This resonance crossing excites their eccentricities to values slightly larger than those currently observed. The small jump in Jupiter's and Saturn's eccentricities destabilizes the motion of Uranus and Neptune, however. The ice giants' orbits become chaotic and start to approach each other. Thus, a short phase of encounters follows the resonance-crossing event. Consequently, both ice giants are scattered outward, onto large eccentricity orbits (e~0.3--0.4) that penetrate deeply into the disk. This destabilizes the full planetesimal disk and disk particles are scattered all over the Solar System. The eccentricities of Uranus and Neptune and - to a lesser extent - of Jupiter and Saturn, are damped in a few My due to the dynamical friction exerted by the planetesimals. Thus, the planets decouple from each other, and the phase of mutual encounters rapidly ends. During and after the eccentricity damping phase, the giant planets continue their radial migration, and eventually reach their final current orbits when most of the disk has been eliminated.

In the framework of this model, the LHB on the giant planets and their satellites is caused by the trans-Neptunian planetesimals as they are dislodged from their primordial disk. Conversely, the bombardment of the terrestrial planets has also a (possibly dominant) contribution by asteroids escaping from the main belt as Jupiter and Saturn crossed their mutual 1:2 mean motion resonance and started to migrate toward their current relative location (Gomes et al, 2005; Strom et al., 2005). In the following section we use this model to quantify the impact rate at Saturn during the LHB.

## 3. Impact rate at Saturn during the LHB

We consider the reference simulation of the Nice model, illustrated in Gomes et al. (2005). From the output of that simulation we get the orbital elements semi-major axis, eccentricity, and inclination $a(t,n), e(t,n), i(t,n)$ as a function of time t for each of the N particles (n=1,..., N), as well as for the giant planets. For every $a(t,n), e(t,n), i(t,n)$ the geometric intrinsic collision probability with Saturn $p_s(t,n)$ and the relative unperturbed velocity $V_\infty(t,n)$ are computed, averaging over all possible orbital configurations occurring during a precession cycle of the orbits. This calculation is accomplished following Wetherill (1967), and using a numerical code implemented by Farinella and Davis (1992) and kindly provided to us. The number of collisions per unit time with Saturn, $P_s(t,n)$, is given by :

$$P_s(t,n) = p_s(t,n)\left(1 + \frac{V^2_{esc\_sat}}{V(t,n)^2_\infty}\right)R^2_s \quad \text{Eq. 1}$$

where $R_s$ is Saturn's radius (set to 58210 km, as an average of Saturn's equatorial and polar radii, Yoder, 1995); $V_{esc\_sat}$ is the escape velocity from Saturn's surface; and $(1+V_{esc\_sat}^2/V_\infty(t,n)^2)$ represents the so-called gravitational focusing factor. For each time t, we then compute the average of $p_s(t,n)$ (called $p_s(t)$) and of $V_\infty(t,n)^2$ (called $V_\infty(t)^2$) over the N particles, weighted by the collision probability $P_s(t,n)$. The cumulative impact probability $pc_s$ at Saturn at time t :

$$pc_s(t) = \int_0^t p_s(t')dt' \quad \text{Eq.2}$$

is shown in Fig. 1 (notice that in the computation of the average, $p\_s(t,n)=0$ for particles that are not Saturn crossers or are no longer active in the simulation). The surge at 850 My corresponds to the trigger of the Late Heavy Bombardment and the achievement of a quasi-stationary value at ~900 My indicates that the overall duration of the LHB at Saturn is about 50 My. Note that 850 My is the time at which the giant planets instability occurred in the reference simulation of Gomes et al., (2005); the lunar basins chronology shows that in the real Solar System the instability may have occurred at 700 My. At the end of the computation, we get $pc_s(10^9$ years$) = 8.41 \times 10^{-15}$. This is the intrinsic collision probability with Saturn per particle in the disk. For comparison, Levison et al., (2000) reports that for today's configuration of giant planets, the total intrinsic probability of ecliptic comets with Saturn is $5.6 \times 10^{-15}$, which is quite close to our estimate above. As for the mean $V_\infty$ our calculation gives 4.69 km/s, whereas for the current Solar System the mean $V_\infty$ of ecliptic comets is ~3 km/s. These comparisons show that the dynamics of comets is weakly dependent on the orbits of the Giant Planets so that, at the time of the LHB, it was similar to that at the current time. Thus the terrific bombardment at the LHB time relative to the current bombardment was simply due to the huge number of planetesimals available at that time, as the primordial massive disk was dispersed.

Given the integrated intrinsic collision probability and relative velocity reported above, the total number of impacts suffered by a satellite of Saturn with radius $r_s$ during the LHB by a planetesimal with radius $r_c$ is computed as (Colwell 1994):

$$N_{LHB}^{Saturn}(r_s, r_c) = N_c(r_c) \cdot 8.41 \cdot 10^{-15} \left(1 + \frac{1}{2}\left(\frac{V_{esc}}{4.69 \text{ km/s}}\right)^2\right)\left(\frac{r_s + r_c}{1 \text{ km}}\right)^2 \quad \textbf{Eq. 3}$$

where $V_{esc}$ is the escape velocity from Saturn's potential well at the satellite orbital distance, and $N_c(r_c)$ is the total number of planetesimals of radius $r_c$ in the primordial disk. Eq.3. is simply interpreted as the number of particles, times the intrinsic impact probability, times the focusing factor at the satellite's location, times the geometrical cross section. The factor of 2 in the denominator of the gravitational focusing factor in Eq.3 accounts for the different geometry of a flux of impactors onto a satellite or ring around a massive planet than onto the planet itself (Morfill et al., 1983, Colwell 1994, Cuzzi and Estrada 1998). Both gravitational focusing factors are approximations to the true enhancement of the impacting flux which depends on the details of the orbits of the impactors. However, our results are quite insensitive to the exact form chosen, and we prefer to use a single formula for all bodies with fixed parameters for simplicity. Note that the number of impacts on Saturn itself is simply obtained by removing the ½ factor from the focusing factor of Eq. 3, by setting $r_s$ to Saturn's radius and $V_{esc}$ to $V_{esc\_sat}$. By doing this, we find that the number of bodies from the primordial Kuiper Belt has a probability of 0.17% to strike Saturn, smaller than the 0.28% probability found by Levison et al., 2000), due to the higher velocity at infinity during the instability of the Nice model.

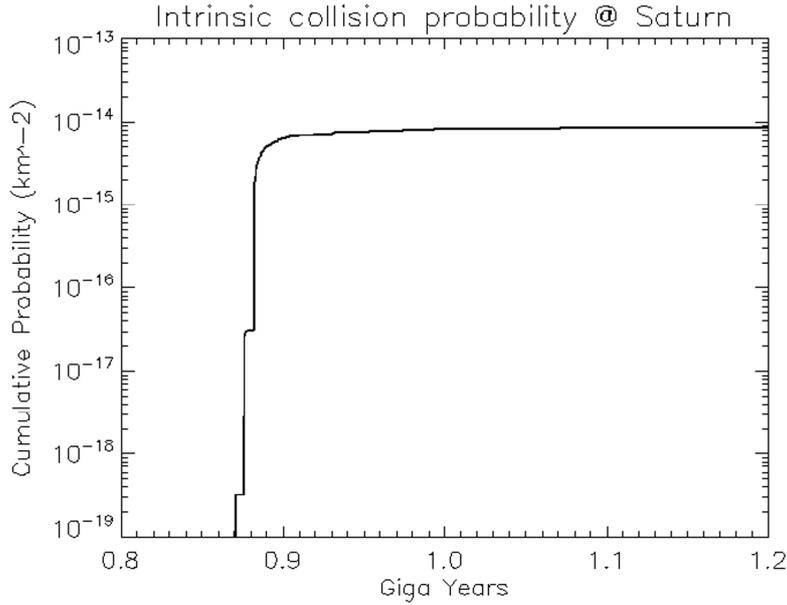

**Figure 1:** Time evolution $p_s(t)$ (km$^{-2}$): the cumulative mean intrinsic impact probability per particle at Saturn.

## 4. Cratering rate on Saturn's satellites

In this section, we assume that bodies that impacted Saturn's system come from the primitive trans-Neptunian disk, in agreement with the LHB scenario of the Nice model (Tsiganis et al., 2005; Gomes et al., 2005). As a first attempt, we assume that the planetesimals in this disk had the size distribution inferred by Charnoz and Morbidelli (2007). The validity of the LHB model with this size distribution is tested against the crater record of Iapetus, which we describe first, because Iapetus is believed to be Saturn's satellite with the oldest surface (see below). Finally, to obtain a better match with the data, we introduce a slight amendment to the size distribution of Charnoz and Morbidelli (2007).

**4.1 Crater record on Iapetus**
In order to benchmark a LHB model for Saturn's system, we need to compare the cumulative impact rate predicted by the model with the crater density on the satellite surfaces that are older than the LHB (i.e. 4.0 Gy).
Unfortunately, the Saturnian system is a dangerous place for satellites: due to its high mass, Saturn's gravitational focusing is effective, and comets falling into Saturn's potential well are accelerated to high velocities before they reach Saturn's satellites. These velocities can be of the order of 30 km/s at a few Saturn radii (about 6 times the impact velocity in the Asteroid Belt). For this reason it has been argued for a long time that the majority of Saturn's regular satellites might not be primordial (Smith et al., 1982; Zahnle et al., 2003). If this is true, the surfaces of these satellites may not have recorded the totality of the bombardment history of the Saturn system. So our reference satellite must be chosen carefully. Iapetus may be a good candidate: its fossil shape (corresponding to a hydrostatic body with a 16-hour rotation period) implies the presence of a strong lithosphere when the satellite was still rotating faster than today (Castillo et al., 2007). Moreover, a strong heat source is required to be active at the time of Iapetus' fast rotation in order to melt its interior; this implies that when the satellite formed, short-lived radioactive elements such as $^{26}$Al had to be present in large abundance, which in turn implies that the satellite formed not later than 5 million years after the formation of the first Solar System solids (Castillo et al., 2007). In addition, Iapetus might have had no endogenic activity since 200 My after its formation (Castillo et al., 2007). Finally, Iapetus has the most heavily cratered surface of all of Saturn's regular satellites (Porco et al., 2005). Thus, all these arguments strongly suggest that the surface age of Iapetus is comparable to the age of Saturn itself. Fortunately, the surface of Iapetus has been observed at high resolution by both Voyager (Smith et al., 1982) and Cassini (Castillo et al., 2007, Giese et al., 2008), so we can take advantage of a large amount of good data. For all these reasons, reproducing the main characteristics of the Iapetus crater record is a good test for constraining any bombardment scenario. We will use three observational tests:

- *The surface density of craters with diameter D larger than 10 km*: it was estimated to be $2.3 \times 10^{-3}$ km$^{-2}$ by Smith et al., (1982), using a power law extrapolation calibrated on the number of larger craters. It has been recently re-evaluated in Neukum et al., (2005) using direct counts from Cassini images of Iapetus' dark terrains. From Fig. 2 in Neukum et al., (2005), we read ~ $3\pm1\times10^{-4}$ km$^{-2}$ craters with diameter D>10 km. We will use the latest published value (Neukum et al., 2005), although we are aware that crater counting is a particularly difficult task that could be very author-dependent.
- *The surface density of craters with diameter D larger than 100 km*: from Fig. 2 of Neukum et al., (2005) we read off $7\pm1\times10^{-6}$ km$^{-2}$.
- *The number of basins larger than 300 km diameter*: Giese et al., (2008) report that about 7 basins with D>300 km are observed on Iapetus' bright side. Crater counting also reveals at least two basins on the dark side (Denk et al., 2008), although this count could be still incomplete. A reasonable range could be that between 10 and 15 D>300 km basins are present on Iapetus' surface (Tilmann Denk, private communication)

These numbers are reported in the second column of Table 1. To convert crater diameter into an impactor size, we use the model by Melosh (1989). We assume that the impact velocity on Iapetus is 7.4 km/s, which is the square root of the quadratic sum of Iapetus' orbital velocity, Iapetus' escape velocity, the escape velocity from Saturn at Iapetus' distance, and the mean velocity at infinity of the projectiles (see section 2). Note that Zahnle et al., (2003) give a mean impact velocity on Iapetus of 6.1 km/s, which is slightly smaller than our value because of the somewhat smaller velocities at infinity that are typical of present-day ecliptic comets encountering Saturn. We assume that both Iapetus and the impactors have densities close to 1000 kg/m$^3$ (Jacobson et al., 2006). The model also depends on the transient crater diameter $D_{tr}$, which may differ from the observed diameter D, when D is larger than a critical diameter $D_t$ (Melosh 1989, Cintala & Grieve 1998). When $D>D_t$ the crater is called "complex" and a correction factor must be applied on D. For Iapetus leading face, the transition diameter from simple to complex crater is $D_t=11\pm3$km (Giese et al., 2008). So D=10 km craters could be considered as simple craters with $D_{tr}=10$ km. Conversely, 100km and 300km diameter craters are complex and Giese et al., (2008) recommend $D_{tr}=D/2.7=37$ and 111 km respectively. With these numbers, the projectile sizes are derived following Melosh (1989): impactors with radii of 290 m, 2.9 km and 6.5 km are needed to form D=10 km, 100 km and 300 km craters on Iapetus, respectively. Therefore, we assume that the crater densities reported above correspond to the cumulative impact rate over Saturn's history for projectiles of these sizes. We use these numbers below to test the validity of our LHB model.

|  | Observation | Required impactor radius (♠) | Simulation with CM07 distribution | Simulation with ISD distribution |
|---|---|---|---|---|
| Surface density of D>10km craters on Iapetus (km$^{-2}$) | 3±1 ×10$^{-4}$ (♣) | 250 m | 3x10$^{-3}$ | 2.0 × 10$^{-4}$ |
| Surface density of D>100km craters on Iapetus (km$^{-2}$) | 7±2 ×10$^{-6}$ (♣) | 2.9 km | 7.7 x 10$^{-6}$ | 6.5× 10$^{-6}$ |
| Number of D>300km basins on Iapetus | 10-15 (♦) | 6.5 km | 6.7 | 13 |

**Table 1**: Resulting cratering rates for CM07 and Iapetus-scaled size distribution of Kuiper Belt Objects (see Fig. 2) during the LHB on Iapetus. (♠) The required impactor radius is computed using the Melosh (1989) cratering model. (♣) From Figure 2 of Neukum et al., (2005). (♦) Extrapolated from D>300km basins found on Iapetus' leading side (Giese et al., 2008).

### 4.2 The initial size distribution of planetesimals in the trans-Neptunian disk.

In the Nice model of the LHB the reservoir of the giant planet impactors is the massive trans-Neptunian disk extending up to ~35 AU (see section 2). This disk is the progenitor of the current Kuiper Belt. So, we can use the size distribution of the Kuiper Belt to infer the size distribution in the primordial disk. It is now well accepted that the current Kuiper Belt shows a deficit of mass relative to its primordial content. The current mass of the Kuiper Belt is estimated to be 0.01 to 0.1 Earth masses (Bernstein et al., 2004, Gladman et al., 2001, Petit et al., 2006), whereas the estimated initial mass is about 10-30 M$_\oplus$ (Stern & Colwell 1997, Kenyon & Bromley 2004; see Morbidelli & Brown 2004 for a review). The mechanisms proposed to explain this mass deficit can be grouped into two broad categories, each of which implies a different initial size distribution: (i) collisional grinding scenarios or (ii) dynamical depletion scenarios. In the collisional grinding scenarios (see e.g.,. Kenyon et al., 2008 for a review) the initial population of big objects (with average radii r >100 km) was never significantly larger than today's population (bodies of such size cannot be destroyed by collisions; Davis and Farinella 1997), and the missing mass was entirely carried by small bodies, which are easy to fragment. From the quantitative point of view, if the collisional grinding scenario is correct, the primordial size distribution at the big size end had to be the same as the current one, culminating with 1-2 Pluto-size bodies (Pluto's diameter ~2400 km). The current steep size distribution (differential power law index q~-4.5, now valid only down to bodies with diameters of ~100 km; Bernstein et al., 2004, Fuentes & Holman 2008) had to be valid down to meter-size bodies (Kenyon and Bromley 2001). However, it has been recently shown (Charnoz & Morbidelli 2007) that such a primordial distribution would raise a problem for both the Scattered Disk (SD hereafter) and the Oort Cloud (OC hereafter) which also originated, like the Kuiper Belt, from the same planetesimal disk. In fact, because of the effective collisional grinding imposed by the steep size distribution, both the SD and the OC would now be deficient in 1-10 km comets by a few orders of magnitude, relative to the current population estimates derived from the flux of Jupiter-family and Long-period comets.

This problem is solved if one assumes that the mass depletion of the Kuiper Belt is due to dynamical processes and not to collisional grinding. In the Nice model, only a tiny fraction of the original disk planetesimals (of order ~0.1%) was implanted into the current belt during the large-eccentricity phase of Neptune's evolution, and survived there up to the present time (Levison et al., 2008). Because dynamical processes are size-independent, the Nice model - as well as any other dynamical depletion model- requires that the initial size distribution in the disk was the same as that observed today in the KB, but multiplied by a size-independent factor (corresponding to the current mass deficit factor, of order of 100 to 1,000). Thus, inspired by the current Kuiper Belt size distribution (Bernstein et al., 2004), Charnoz & Morbidelli (2007) assumed:

$$\begin{cases} \dfrac{dN}{dr} \propto r^{-3.5} & \text{for } r < 100\text{km} \\ \dfrac{dN}{dr} \propto r^{-4.5} & \text{for } r > 100\text{km} \end{cases} \quad \textbf{Eq. 4}$$

and showed that in this case both the Scattered Disk and the Oort Cloud would contain a number of 1-10 km objects consistent with the population estimates obtained from the observed fluxes of comets.

In order to have ~30$M_\oplus$ of planetesimals in the outer Solar System, with the size distribution of Eq. 4, about 300 Pluto-sized bodies had to be present initially. This distribution is called CM07 hereafter (figure 2, dotted line). Using the intrinsic collision rates derived in section 2, and assuming that comets with r>290 m produce craters with D>10 km, the resulting surface density of craters with D>10 km would be 0.0023/km$^2$. This compares well with the crater density estimated by Smith et al. (1982). However, as we said in sect. 3.1, the crater density computed in Neukum et al., (2005) might be more correct, and it is almost an order of magnitude smaller (see Table 1, fourth column). The resulting surface density of D>100km craters is 7.7×10$^{-6}$/km$^2$, in very good agreement with Neukum et al., (2005). Concerning basins with D>300 km, our model predicts a number of ~6.7, in reasonable agreement with the estimates (between 10 and 15 in total, see section 4.1).

This essentially validates, at least at the order of magnitude level, the Nice model of the LHB with the disk planetesimal size distribution of Eq. 4. Below, we discuss a slight amendment of the size distribution that allows us to achieve a better match with the Neukum et al. (2005) count of D=10 km craters, and with the number of basins on Iapetus' surface

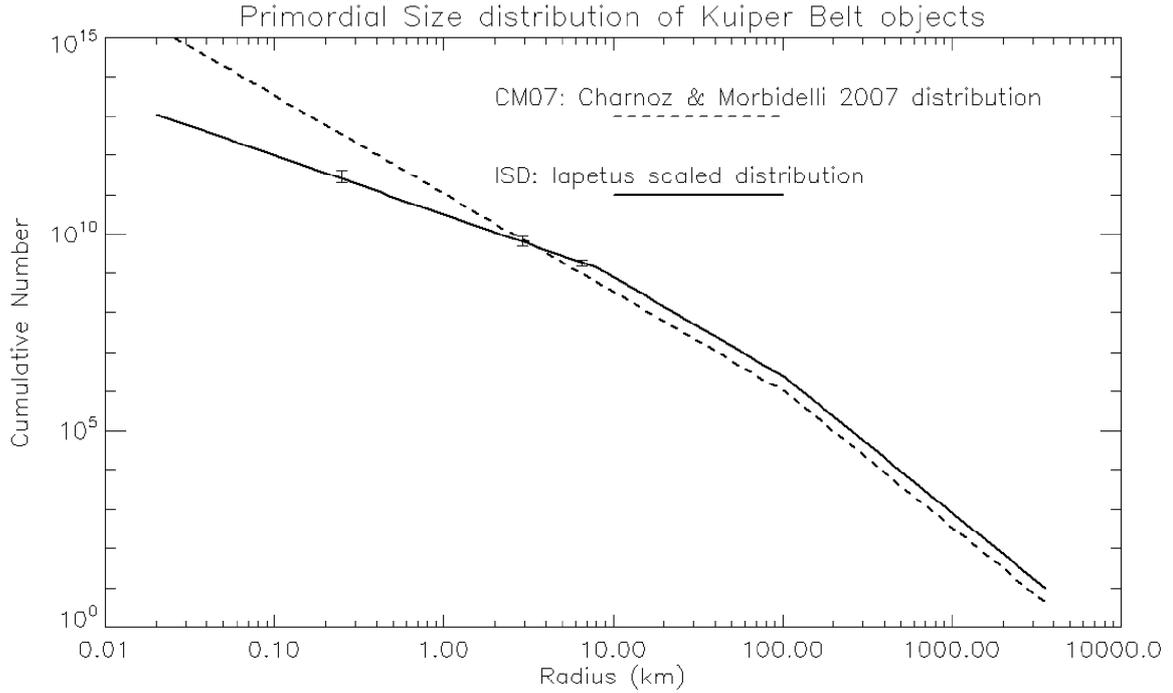

**Figure 2:** The two primordial size distributions of the trans-neptunian disk considered in the present paper. The CM07 distribution (taken from Charnoz & Morbidelli 2007) was originally proposed as a good match for producing today's Kuiper Belt, Scattered Disk, and Oort Cloud. The Iapetus-scaled distribution (ISD) is a modification of CM07 in order to account for (i) craters observed on Iapetus and (ii) the lack of comets with D<15 km in the inner Solar System (Zahnle et al. 2003). Error bars located at R=0.29 km, 2.9 and 6.5 km show the acceptable range of values for reproducing the abundance of craters on Iapetus, as reported in the second column of Table 1 (see section 4.2).

### 4.3 An Iapetus-scaled size distribution

As we have seen, the Nice model of the LHB, with the size distribution of Eq. 4, might overestimate the number of craters caused by projectiles ~300 m in radius. The crater records on Jupiter's satellites indeed suggest that the population of comets with diameter <20 km have a substantial deficit compared to a power law extrapolation with an exponent q=-3.5 (as in Eq. 4), calibrated on the number of large bodies (Zahnle et al., 2003). In fact the exact location of the knee in the crater size-distribution is uncertain and could be anywhere between 1 and 20 km diameter (Lowry et al., 2003; Zahnle et al., 2003). This deficit of small size objects may be due to cometary disruption in the inner Solar System, or could be primordial, reflecting the size distribution of bodies in the Scattered Disk (see e.g., Whitman et al., 2006, for a discussion). Several authors agree that the cumulative power-law index of the size distribution of comets with diameters less than 15 or 20 km is in the range -1.4 to -1.6 (Donnison 1986; Lowry et al., 2003; Zahnle et al., 2003; Whitman et al., 2006; Fernández and Morbidelli 2006). So a cumulative exponent of -1.5 (corresponding to a -2.5 differential index) will be used as an intermediate value, for comets with diameter smaller than 15 km.

After testing different possibilities, and assuming that the original distribution was a combination of simple power laws, we adopt the following "Iapetus Scaled Distribution" (ISD hereafter):

$$\begin{cases} \dfrac{dN}{dr} \propto r^{-4.5} & \text{for } r > 100\text{km} \\ \dfrac{dN}{dr} \propto r^{-3.5} & \text{for } 7.5\text{km} < r < 100\text{km} \\ \dfrac{dN}{dr} \propto r^{-2.5} & \text{for } r < 7.5\text{km} \end{cases} \quad \text{Eq. 5}$$

with 800 Pluto-sized bodies (see Fig. 2, distribution in bold line). With this size distribution, the number of D>300 km basins on Iapetus increases to 13, in good agreement with what is observed (between 10 and 15 basins on Iapetus) and the density of craters with D>10 km and D>100km shifts to $2.0\times10^{-4}$ km$^{-2}$ and $6.5\times10^{-6}$ km$^{-2}$ respectively, in very good agreement with the Neukum et al. estimates (section 4.1, and Table 1 fifth column). Moreover, we stress that the assumption of the existence of ~800 Pluto-size objects (instead of 300) is also in better agreement with the Nice model (which implies ~1,000 Plutos in the disk, see Levison et al., 2008).

Notice that the size distribution in Charnoz and Morbidelli (2007) (see Eq. 4) was derived from considerations on the number of 500-m comets in the Oort cloud and in the scattered disk. With the distribution of Eq. 5, the number of 500-m objects is decreased by a factor of 7.5, relative to the distribution of Eq. 4. On the other hand, the shallower distribution of Eq. 5 would give less collisional grinding than estimated in Charnoz and Morbidelli, so that the final numbers of 500-m comets in the Oort cloud and in the SD would probably agree within a factor of a few with the values in CM07. Moreover, if the shallow distribution of cometary nuclei is due to their physical disruption before they reach Saturn's system, Eq. 5 should be interpreted as the *size distribution of the projectiles on Saturn's system*, and not as the original size distribution in the planetesimal disk. The latter can still have a slope with q~-3.5, as in Eq. 4 and in Charnoz and Morbidelli (2007).

Using a size distribution with three slopes may seem artificial, at first sight, just to match the Iapetus' crater density. However, we remind the reader that this size distribution comes from self-consistent considerations with the formation of the Kuiper Belt, Scattered Disk and Oort Cloud, which imposes the two slopes beyond R=7.5km (Charnoz and Morbidelli 2007). The slope below 7.5 km is severely constrained by the crater record on Iapetus (see Fig.2). This is illustrated by the error bars displayed in Fig.2 (error bars are derived from the uncertainties reported in the second column of Table 1). We note that these are very narrow ranges, and that the three constraints on Iapetus' craters are met by a same slope of our size distribution, suggesting this is quite robust in the frame of our assumptions :(1) KBOs are the primary source of impactors on Iapetus and (2) the ½ term in front of the focusing factor in Eq. 3 is correct (see section 3). In the case the ½ factor is wrong (an should be replaced by 1), we can still get a very good fit to the data using the size distribution of Eq.5, but scaling it to 600 Pluto-sized bodies (rather than 800), and the results of the paper remain unchanged.

# 5. Implications for the origin of Saturn's ring system

On the basis of the Nice model for the LHB and the projectile size distribution given in Eq. 5, we now revisit the two main scenarios for the formation of Saturn's rings.

### 5.1 Scenario 1: Tidally disrupted comets
As suggested in Dones (1991) and in Dones et al., (2007), a possible scenario for the formation of Saturn's rings is the tidal disruption of one, or several, comets passing inside Saturn's Roche Zone on a hyperbolic orbit with a low asymptotic velocity ($V_\infty$). During the tidal disruption event, the orbital energy is spread among the fragments. As a consequence, a fraction of the comet's fragments can be captured on bound orbits around the planet, while the rest escape from the sphere of influence of the planet (Dones 1991). Then, the collisions among the captured fragments circularize the fragments' orbits, reduce the semi-major axes, and grind the fragments down to smaller sizes. Subsequently, dissipation in physical collisions flattens the particle swarm into a thin disk, forming a ring system of centimeter-to-meter-sized particles. In this scenario the incoming comets must pass close enough to Saturn and with a sufficiently low relative velocity at infinity. On the basis of simple energetic considerations, the fraction of cometary material that is captured onto a bound orbit with initial apocenter distance smaller than $R_{stab}$ is (Dones, 1991):

$$f = \frac{0.9\Delta E - GM_s/R_{stab} - 0.5V_\infty^2}{1.8\Delta E} \quad \textbf{Eq. 6}$$

where $\Delta E = GM_s r_c/q^2$ stands for the difference in orbital energy across a comet with radius $r_c$, whose closest approach distance to the center of Saturn is q. The fraction of mass that is captured depends on (i) the distance of closest approach, (ii) the comet's radius, and (iii) on the velocity at infinity. $R_{stab}$ must be smaller than Saturn's Hill radius. To estimate the total implanted mass, we first compute the distribution of $V_\infty$ of passing comets from the numerical simulations of the Nice model. This distribution is shown in cumulative form in Fig. 3. Then, for each size bin in the comet size distribution, we compute two arrays in (q,$V_\infty$):

- $N(q, V_\infty)$: the number of comets approaching the planet, which can be a non-integer quantity, as it is obtained by multiplying the encounter probability $P(q, V_\infty)$ by the number of comets in the size distribution of Eq. 5. This quantity is the same for comets of all sizes.
- $f(q, V_\infty, r_c)$: The mass fraction of a comet (with radius $r_c$, passing at pericenter q with a velocity $V_\infty$) that is captured onto a planetocentric orbit, as given by Eq. 6.

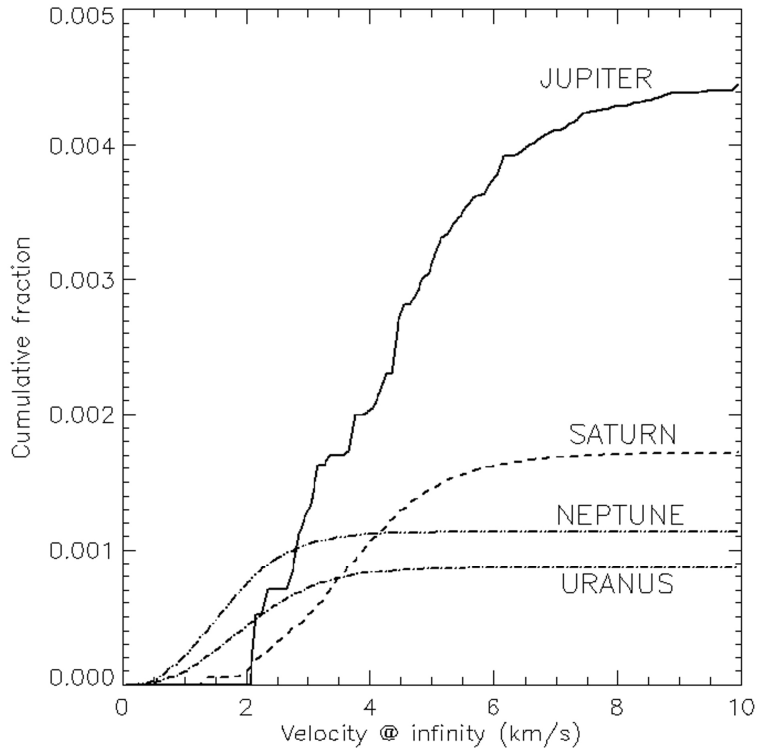

**Figure 3**: Fraction of particles starting in the primordial trans-neptunian disk impacting the giant-planet's surface with velocity at infinity smaller than a given value.

The pericenter distance q was binned in 100 bins ranging from 1Rs to 2.5Rs and $V_\infty$ was binned in 100 bins between 0 and 10 km/s. The total implanted mass is obtained by summing all bins of $F \times N \times m_{comet}$ (with the comet mass $4/3 \pi r_c^3 \times 1000$ kg/m$^3$). Because it turns out that most of the implanted mass is provided by the most massive bodies (objects with radii from 500 to 2,000 km; see Fig. 4) during rare events (i.e. with the total expected number of events N<1), we decided to use a Monte Carlo simulation, in the spirit of Dones (1991), in which we determine the number of events using a random number generator and considering that the fractional part of N is the probability for the last event to happen. Obviously the use of the Monte Carlo approach is significant only for the cases with N ≤ 1, but – as we said above - these rare events are those that carry most of the mass (lower panel of Fig.4). In total we did 100 Monte Carlo simulations for each planet and computed the mean and median mass implanted as a function of $R_{stab}$ (see Table 2).

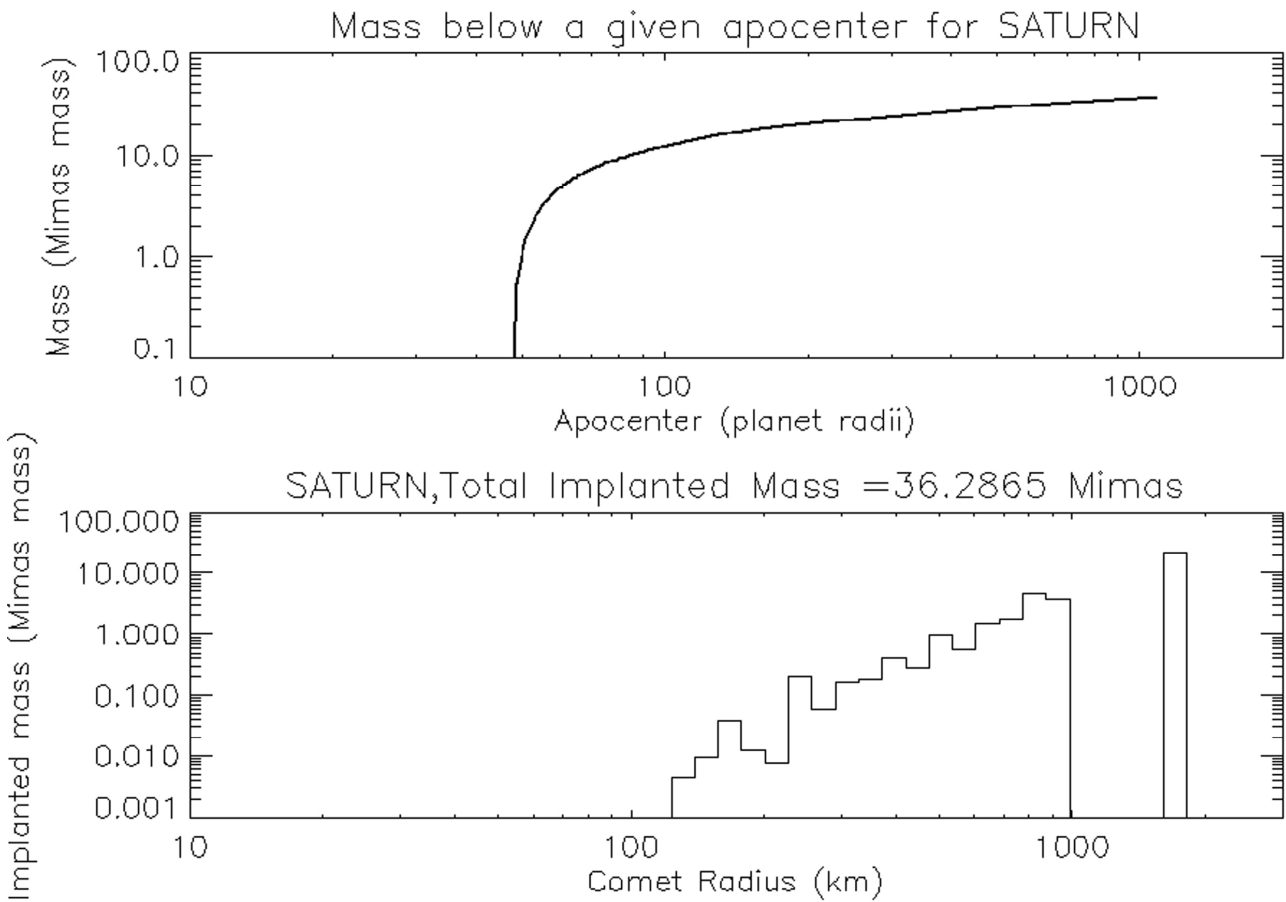

**Figure 4:** Mass injected into Saturn's Hill sphere via tidal disruption of comets. Top: Cumulative distribution of debris' apocenters. Note the pericenters are between 1 and 2 planetary radii. Bottom: Mass injected as a function of the comet size. The ISD distribution of impactors was used here (see section 4.3 and Fig. 2).

[ PUT TABLE 2 HERE ]

The fact that the mean and median values appear very different is precisely an indication that most of the mass is carried in events with a probability to happen smaller than unity. This is also illustrated by the large dispersion of the values. We find that the mass captured around Saturn by tidal splitting of passing comets is big, of order of tens of times of the mass of Mimas, at least for a large initial apocenter distance $R_{stab}$. We repeated the calculations for the other giant planets and, ironically, Saturn turns out to be the planet that captures the smallest amount of material, whereas Neptune is the planet that captures the largest amount. This is a consequence of several factors: Neptune is the densest and the most distant of all the giant planets, so that the ratio between its physical radius and its Hill radius is the smallest (for instance, Neptune's Hill radius is about 4700 planetary radii whereas Saturn's Hill radius is only 1100 planetary radii). Note also that Neptune's Hill radius is about twice as large as Saturn's Hill radius in physical units. All this results in a larger efficiency of capture: indeed, a little algebra with Eq.6 shows that f increases with smaller values of q (comparable to the planet's radius) and larger values of $R_{stab}$ (comparable to the planet's Hill sphere). In addition, because Neptune is the closest planet to the outer planetesimal disk (at least from the time of the triggering of the LHB) the number of comets passing close to Neptune is the largest among all four giant planets. In conclusion, the tidally disrupted comet scenario seems to work, in principle, during the LHB, but it has a big problem: it predicts that all giant planets should have acquired rings at least as massive as Saturn's. Unless the LHB model is totally wrong in predicting the relative fluxes of comets in the vicinity of the giant planets (this seems unlikely because, as we remarked in section 2, the dynamics of the comets in the Nice model is very similar to the current one, so that it is not very sensitive to the exact planetary evolution; in other words: how could comets encounter Saturn but not Jupiter, Uranus and Neptune?), something must make the tidal disruption scenario much less efficient than it seems at a first examination.

A possible explanation for a lower efficiency is that tidally disrupted comets should be roughly half on prograde orbits and half on retrograde orbits with respect to the planet (Zahnle et al., 1998, Levison et al., 2000). Thus the total angular momentum of the debris of all comets taken together should be close to zero. Consequently, collisional damping – which changes a and e but preserves the total angular momentum - should cause the fall of most of the material onto the planet. To test this effect, the implanted total orbital momentum carried by the trapped material was also computed (Table 2). It is defined as:

$$J_{tot} = \sum_{fragments:f} r \times m_f \sqrt{GM_s a_f (1-e_f^2)} \quad \text{Eq. 7}$$

where $m_f$, $a_f$ and $e_f$ are the total mass, semi-major axis and eccentricity of the ensemble of objects captured from the splitting of one comet, and r is set to be +1 or -1 at random with equal probabilities. Statistics on the distribution of implanted angular momentum are reported in Table 2. Positive and negative angular momentum mean prograde and retrograde rotation, respectively. We see again that large variations are possible and that on average, as for the total implanted mass, the final angular momentum budget is dominated by a few rare events involving massive objects because: (i) the angular momentum is directly proportional to the mass of the incoming body and (ii) the biggest comets implant their fragments on orbits with lower values of eccentricity, which in turn have a larger angular momentum, increasing the relative weight of the largest incoming bodies. We see also that the standard deviation of the implanted angular momentum exceeds by far the its average value. It is then impossible to predict a-priori what the sign of the total angular momentum of the resulting ring system will be. If such an explanation was valid to explain the origin of planetary rings in our Solar System, we should observe some rings orbiting in the prograde direction, and others in the retrograde direction, which is obviously not the case. So, this explanation does not seem to work.

Another possible explanation is suggested by Table 2 (two last rows) and highlighted in the top panel of Fig. 5. For all giant planets, the vast majority of the material captured from the tidal disruption events has an initial apocenter distance larger than several hundred times the planetary radius, namely orbital eccentricities larger than 0.99. Material on such high eccentricity orbits may be very unstable and small perturbations (collisions or gravitational perturbations) may eject it from Saturn's sphere of influence. The distribution of apocenter distances (from which we derived the distribution of eccentricities) is computed simply by assuming a uniform distribution of orbital energies for the fragments (Eq. 6 is based on this assumption; see Dones 1991), and varying the value of $R_{stab}$ acting like the value of apocenter. The apocenter distribution falls abruptly within a few tens of planetary radii. In fact, a little algebra shows that the minimum comet radius for implanting material with apocenter smaller than $R_{stab}$ is $R_{min} \approx q^2/R_{stab}$ (assuming $V_\infty=0$, the most optimistic case, see Dones 1991). For Saturn we obtain $R_{min}$= 62 km, 1200 km, and 3016 km for $R_{stab}$=1000, 50 and 20 $R_s$, respectively. Since KBOs with radii larger than 2500 km are very rare in our impactor distribution (Eq. 5), no mass can be implanted on orbits with apocenter distances below 20 Saturn radii, resulting in eccentricities larger than 0.9 (since the pericenter is around 1 planetary radius). So, given that most of the captured mass consists of fragments on nearly-parabolic orbits, it is difficult to say which fraction of this material could survive tiny perturbations (solar perturbations, perturbations from passing planetesimals or scattering from the regular satellites of the planets may eject the material from the Saturnian system). Also, a fraction of the captured material may collide with the satellites of the planet.

A third possible explanation is that the Dones (1991) model of tidal capture assumes that the incoming body breaks into an infinite number of particles, each on its own keplerian orbit. Models of tidal disruptions can be very different, depending on the material strength and structure (e.g., rubble pile or solid body) and for the moment we have no idea of the internal structure of primordial bodies of the large Kuiper Belt objects. However, some studies of tidal splitting (Dobrovolskis 1990, Davidsson 1999, Holsapple & Michel 2008) suggest that one main fracture fault appears in the body structure and thus the body could split into a few big objects rather than into a huge number of small particles. These big objects could be more likely to return to infinity as they contain a significant fraction of the total energy of the incoming progenitor, which is positive.

A final possibility is that the rings of Uranus, Neptune and Jupiter, are destroyed by some processes in less than 4.5 Gy, which would explain why they are so tenuous today. However, this would not explain why they all four rotate in the prograde direction, given that numerical simulation suggest near equal probabilities for prograde or retrograde rings (see above, and Table 2). However, a substantial mass delivery to the four giant planets seems unavoidable, and its consequence are not still understood. Because of conservation of angular momentum and collisional evolution, a fraction of this exogenic material may end in the planet's Roche Zone, but what fraction? On which timescale? On what orbits? For the moment we do not know and this deserves to be studied in the future.

In summary, it is likely that tidal disruption is only part of the story, but not the full story in the formation of rings, and that other, low-efficiency factors should enter into the game. We will return to this in the discussion section.

## 5.2 Scenario 2: Collisional destruction of a primordial satellite

Given that the estimated mass of Saturn's rings (e.g. Esposito et al., 1983, 2007 Stewart et al., 2007) is typical of Saturn's main satellites (of the order of Mimas' mass), it has been proposed that Saturn's rings could have been produced by the break-up of an ancient satellite, destroyed by a passing comet (Pollack et al., 1973, Pollack 1975, Harris 1984). Before estimating the impact probabilities, we must comment on some intrinsic difficulties of this scenario. One of the most difficult points is: how to bring a satellite with a radius of order of 200–400 km inside Saturn's Roche zone? Another point is: could a satellite survive and remain in the Roche Zone until the LHB begins, 700 My after planet formation? We try to address these two questions below.

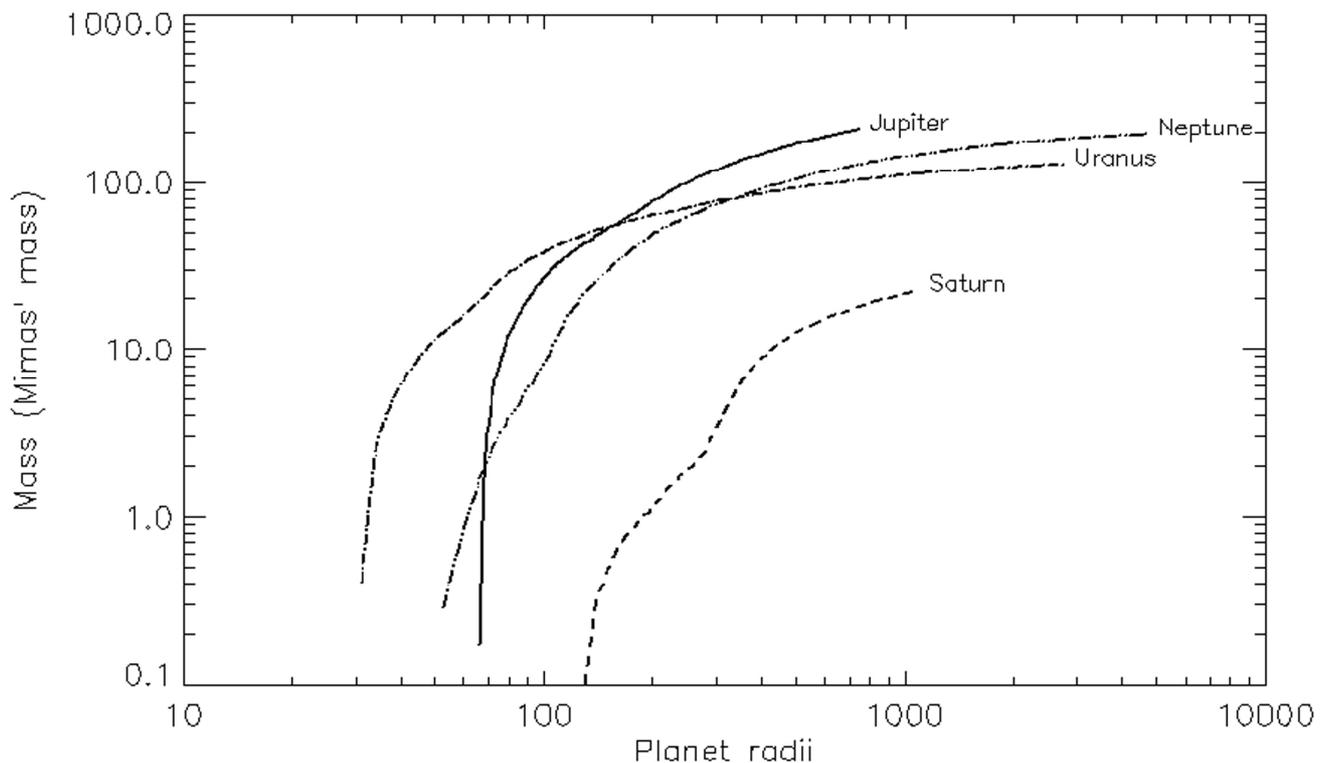

**Figure 5**: Distribution of apocenters (total mass of cometary debris with orbit' apocenters are below a given distance) for material injected into the Hill Sphere of the four giant planets. Masses are in units of Mimas. For each planet, 10 different LHB events were generated (using a Monte-Carlo procedure), and averaged to produce this plot.

### 5.2.1 Implanting a satellite in Saturn's Roche Zone.
In recent years, it has been proposed that regular satellites of the giant planets form in "gas-starved" circumplanetary disks (Canup & Ward 2002, 2006). This model is appealing for two reasons: (i) it explains why the total mass of a satellite system scales approximately linearly with the mass of the central planet, with a typical mass ratio of ~$10^{-4}$ (corresponding to ~ 1500 Mimas' masses for Saturn's system) and (ii) it predicts a relatively late formation of the surviving satellites, consistent with the observation that Jupiter's satellite Callisto appears not to be differentiated. In the Canup and Ward scenario, the satellites suffer an inward type-I migration through the circumplanetary disk, and all satellites that form early are lost by collision onto the planet. When the circumplanetary nebula disappears, type-I migration stops and the surviving satellites remain frozen on the orbits that they have achieved at the time. In this context, it is not unlikely that a satellite is eventually found inside the planet's Roche Zone, brought there by type-I migration but not driven all the way into the planet because of the timely disappearance of the disk (see e.g., Fig. 1 in the Supplementary Online Material of Canup & Ward 2006).

### 5.2.2 Survival in Saturn's Roche Zone
Once in Saturn's Roche Zone at distance $a_0$, and after the dissipation of Saturn's sub-nebula, the satellite's orbit and internal structure would be affected by the tidal interaction with the planet. Given that the dissipation of the nebula happened within ~10 million years, while the LHB happened about 700 million years later, it is important to address what may happen to a satellite Saturn's Roche Zone during this long time interval. We comment below on two critical aspects: (1) tidal splitting and (2) tidally driven migration.

### 5.2.2.1 The tidal splitting of the satellite
A popular - but simplistic - idea is that if a satellite enters a planet's Roche Zone, it is rapidly destroyed because of tidal stresses, so that its fragments should eventually form a ring. However, simple arguments may refute such a scenario: on the one hand, if the satellite were ground to small particles, with sizes of order of mm to cm, as soon as it migrated into the Roche Zone, the particles would have suffered an intense aerodynamic drag due to the gas that still had to be present in order to drive the migration of the satellite. Consequently, solid material would have been rapidly evacuated into the planet. In summary, a debris disk resulting from a satellite destruction created prior to the dissipation of the nebula would not survive. On the other hand, studies of tidal splitting show that a spherical satellite, assembled outside the Roche Limit and brought close to the planet, would survive quite deep inside the planet's classical Roche Zone, mainly because it is solid and not liquid (Holsapple and Michel 2006). In a model relevant for spheres on circular and spin-locked orbits, Dobrolovskis (1990) showed that the fracture regime depends on the relative values of the tensile strength, T, to the satellite's internal pressure $P_0$ (Davidsson 1999):

$$P_0 = \frac{2}{3} G \pi \rho_s^2 R_s^2 \quad \text{Eq. 8}$$

For a satellite with $R_s$~250 km and $\rho_p$=1000 kg/m$^3$, we get $P_0$~ 8.7x10$^6$ Pa. As a rule of thumb, we assume T~3x10$^7$ Pa, halfway between pure rock and pure ice (Dobrovolskis 1990). Thus, depending on the fracture regime, the distance for tidal splitting could be well inside the Roche Zone – between 0.6 to 1.3 planetary radii (equations (5) to (11) of Davidsson 1999), i.e., below 76000 km in the case of Saturn. We also note that another similar model suggests that a 100 km radius satellite can survive undisrupted at 100,000 km from Saturn's center, i.e., where the B ring currently lies (Goldreich & Tremaine 1982). So it seems that a satellite with dimensions comparable to Mimas may survive deep inside the classical Roche Zone: it could be elongated in Saturn's direction, but not broken. Moreover, even if tidal splitting occurred, the process might not grind a satellite all the way into small particles; instead, big chunks of material, with stronger tensile strength, might survive undestroyed (Keith Holsapple, private communication).

So it seems unavoidable that a destruction process different from tidal splitting is required to break the satellite into small particles, even if the satellite was originally in the Roche Zone. A cometary impact during the LHB could be a solution, as discussed in section **5.3**. We note also that satellites, some with radii approaching 100 km, orbiting in the Roche Zone of their hosting planet are not so rare in the Solar System, strengthening the above arguments: Phobos orbits deep inside Mars' Roche zone, Amalthea orbits in Jupiter's Roche zone; Pan and Daphnis orbit in Saturn's Roche Zone; and Naiad, Thalassa, Despina, Galatea, and Larissa orbit in Neptune's Roche Zone. We come back to this point in section 6.2.2.

**5.2.2.2 Orbital evolution inside the Roche Zone.**

When a satellite is close to a planet, it raises a tidal bulge which, in turn, induces an orbital migration of the satellite itself. The migration direction depends on the satellite's position, $a_0$, relative to the Synchronous orbit $a_s$ ($a_s \sim 112,000$ km for Saturn, assuming a planetary rotation period of ~10 hours, 40 minutes). If $a_0 < a_s$ then the satellite's semi-major axis, $a(t)$, decays due to transfer of angular momentum to the planet. Conversely, for $a_0 > a_s$ $a(t)$ increases due to angular momentum transfer to the satellite. The time evolution of $a(t)$ is given by (Murray & Dermott 1999):

$$\frac{da}{dt} = sign(a - a_s) \frac{3 k_{2p} m_s G^{1/2} R_p^5}{Q_p m_p^{1/2} a^{11/2}} \quad \textbf{Eq.9}$$

where $m_s$, $m_p$, $R_p$, and $G$, stand for the satellite's mass, the planet's mass, the planet's radius, and the gravitational constant, respectively; $Q_p$ and $k_{2p}$ stand for the dissipation factor and the Love number of the planet. For Saturn $k_{2P}$ is reasonably well known, because it is linked directly to the $J_2$ gravitational moment and spin frequency of the planet ($k_{2P} \sim 0.3$ for Saturn; Dermott et al., 1988). However, the value of $Q_p$ that describes all dissipative processes in the planet's interior is very uncertain. Several indirect estimates exist, but they are still badly constrained: Dermott et al., (1988) suggest $Q_p > 1.6 \times 10^4$ because of the proximity of Mimas to Saturn. More recently, Castillo et al., (2008) suggests that $2 \times 10^{-6} < k_{2P}/Q_p < 3.5 \times 10^{-5}$, yielding $8.6 \times 10^3 < Q_p < 1.5 \times 10^5$ for $k_{2P} \sim 0.3$. For comparison, Jupiter's $Q_p$ is typically considered to be $\sim 10^5$, but a value as high as $10^6$ cannot be excluded (Peale 2003). Moreover, the value of $Q_p$ may have varied by orders of magnitude over the age of the Solar System (Wu, 2005).

We performed several integrations of Eq. 9 varying the satellite's mass and the value of $Q_p$ (Fig. 6). The satellite was started close to Saturn's synchronous orbit to allow a maximum residence time in the Roche Zone (so that a lower bound for $Q_p$ is derived): we assume $a(t=0)$ to be either 115,000 km or 108,000 km (recall that $a_s \sim 112\,000$ km). For a satellite more massive than Mimas and starting below the synchronous orbit, and for any value of $Q_p < 3 \times 10^5$, the orbital decay is so rapid that the satellite would hit the planet before the LHB. Conversely, if a satellite starts exterior, but close, to the synchronous orbit, it may survive inside the Roche Zone up to the LHB epoch, for a wide range of $Q_p$ values. More precisely, for masses of 1, 3 and 5 Mimas' masses, $Q_p$ must be larger than $3 \times 10^4$, $8 \times 10^4$ and $3 \times 10^5$, respectively, in order for the satellite to survive in the Roche Zone up to the LHB epoch (Fig. 7). All these values are within acceptable ranges of our actual knowledge of Saturn's ring mass and $Q_p$. We also note that the B ring, which is the most massive of Saturn's rings, lies precisely around the Synchronous orbit, which may suggest that the putative satellite was destroyed there.
In conclusion, it is possible that a satellite of a few Mimas masses can survive for 700 My in Saturn's Roche Zone, provided it was close to, but outside of, the Synchronous orbit when Saturn's sub-nebula disappeared, and that $Q_p \geq 10^5$ in average during the first 700My of Saturn's history.

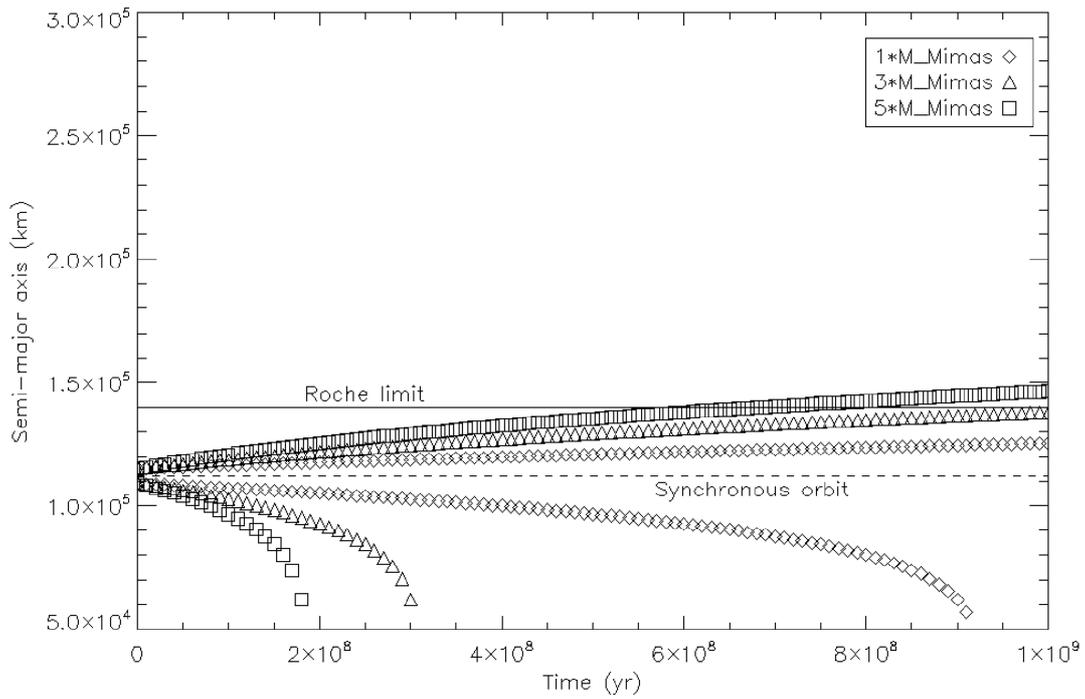

**Figure 6 :** Orbital evolution of satellites, under Saturn's tides, with different masses and assuming $Q_p=10^5$. Satellites exterior to synchronous orbit start with $a_0=115000$ km, and satellites interior to synchronous orbit start with $a_0=108000$ km.

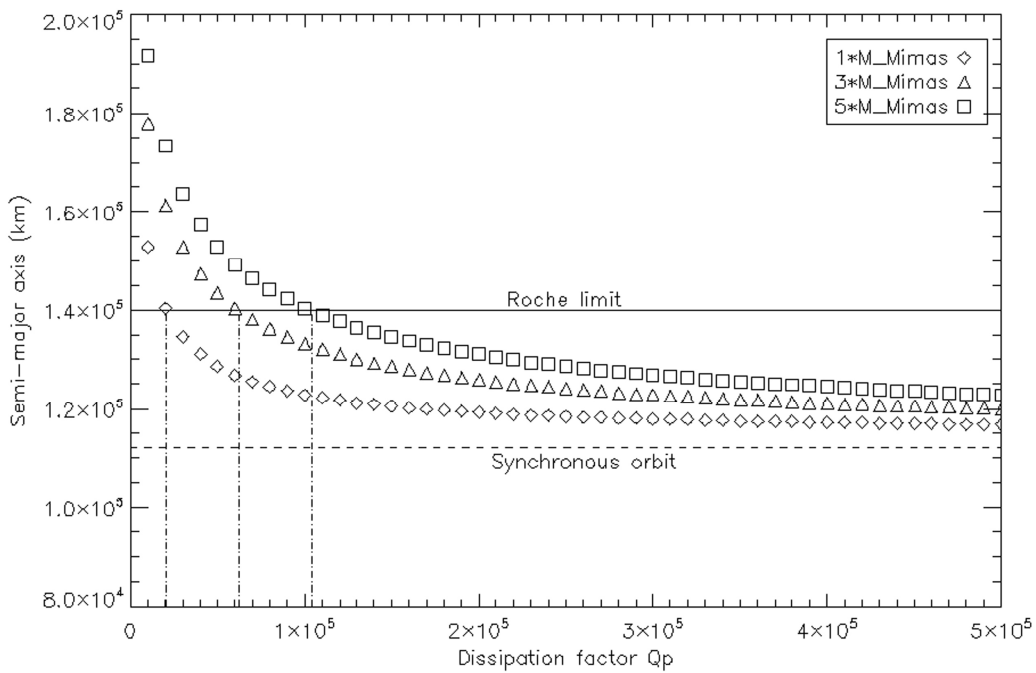

**Figure 7:** Final semi-major axis, as a function of Saturn's dissipation factor $Q_p$, after 700 My of tidal evolution for different satellite masses, all starting at $a_0=115,000$ km. Dot-dashed line designates the minimum $Q_p$ value for a satellite needed to stay in Saturn's Roche Zone for 700 My for the three different satellite masses shown.

**5.2.3 Number of destructions and survival probabilities of Saturn's satellites**

We now assume that a satellite is in Saturn's Roche Zone at the time of the LHB and we quantify its probability of destruction during the LHB. The typical impact velocity, $V_i$, of a satellite on a circular orbit with a comet coming from infinity is $V_i \sim (3GM_p/a+V_\infty^2)^{1/2}$ (neglecting the satellite's gravitational focusing; the factor 3 in front of $GM_p/a$ comes from the gravitational focusing of the planet plus the squared orbital velocity of the satellite, Lissauer et al., 1988), yielding $V_i \sim 32$ km/s at ~110,000 km. This is about 6 times the impact velocity in the Asteroid Belt. Thus, collisions so close to Saturn are particularly destructive. We now need to estimate the impactor size required to destroy the satellite. Unfortunately, to our knowledge, no experiment or numerical model of collisions at such a high impact velocity is available in the literature, so we extrapolate from results obtained for icy bodies impacting at 3 km/s, as given in Benz & Asphaug (1999) using hydrocode simulations. Comparison of Q* at different impact velocities shows only small variations of Q* (for V ranging from 0.5 to 3 km/s), so values of Q* from Benz & Asphaug (1999) could be considered about the right order of magnitude at ~30km/s (M.J. Burchell and A. Lightwing, private communication). The mass ratio, f, of the largest fragment to parent body is approximated by (Benz & Asphaug 1999, Eq. 8):

$$f = -s\left(\frac{Q}{Q^*} - 1\right) + 0.5 \qquad \textbf{Eq. 10}$$

where Q is the specific impact energy (the kinetic energy per unit mass in the center-of-mass frame) and the $Q^*$ is the critical energy for destruction. The value of s is ~0.6 for ice (Benz & Asphaug 1999). $Q^*$ is taken from Eq. 6 of Benz & Asphaug (1999). We note that f is positive only for Q/Q*<11/6 and so, one may wonder the validity of this equation. In the regime of f~0.5, which we use here, Eq. 10 has the advantage of being a very good fit to the Benz & Asphaug (1999) simulations. Other models could be used like the Fujiwara et al., (1977) disruption threshold (f $\propto Q^{-1.24}$), but they do not match hydrocode simulations very well. So we think that Eq. 10 is good enough for the present paper. Strictly speaking, a catastrophic collision is defined as one with Q such that f≤0.5. Using f=0.5 in Eq. 10 (equivalent to Q=Q*) and assuming an impact velocity of 32 km/s, we find that, for a progenitor mass of 1, 3 and 5 Mimas masses (⇔200, 300, 350 km radius), the projectile must be ~17, 31 and 39 km respectively for catastrophic disruption. (assuming that the comet and satellite's densities are both 1000 kg/m$^3$). The number of such events is computed using Eq. 3, using the impactor size distribution of Eq. 5 (see section 4.2). The results are displayed in Fig. 8. A ring progenitor with, 1, 3 and 5 Mimas mass would suffer ~2.6, 1.5 and 1 catastrophic impacts. So these results suggest that if a satellite had been located inside Saturn's Roche Zone, and with mass comparable to or a few times that of Mimas, it could have been destroyed during the LHB with a substantial probability. The fragments would have been scattered around the progenitor's orbit inside Saturn's Roche Zone, and could have not re-accreted, because of the combination of (i) the planet's strong tides and (ii) the intense perturbation by the close and massive core stirring up the debris disk. This would have led to the formation of a massive disk, with a mass equal, or comparable to, the parent body's mass.

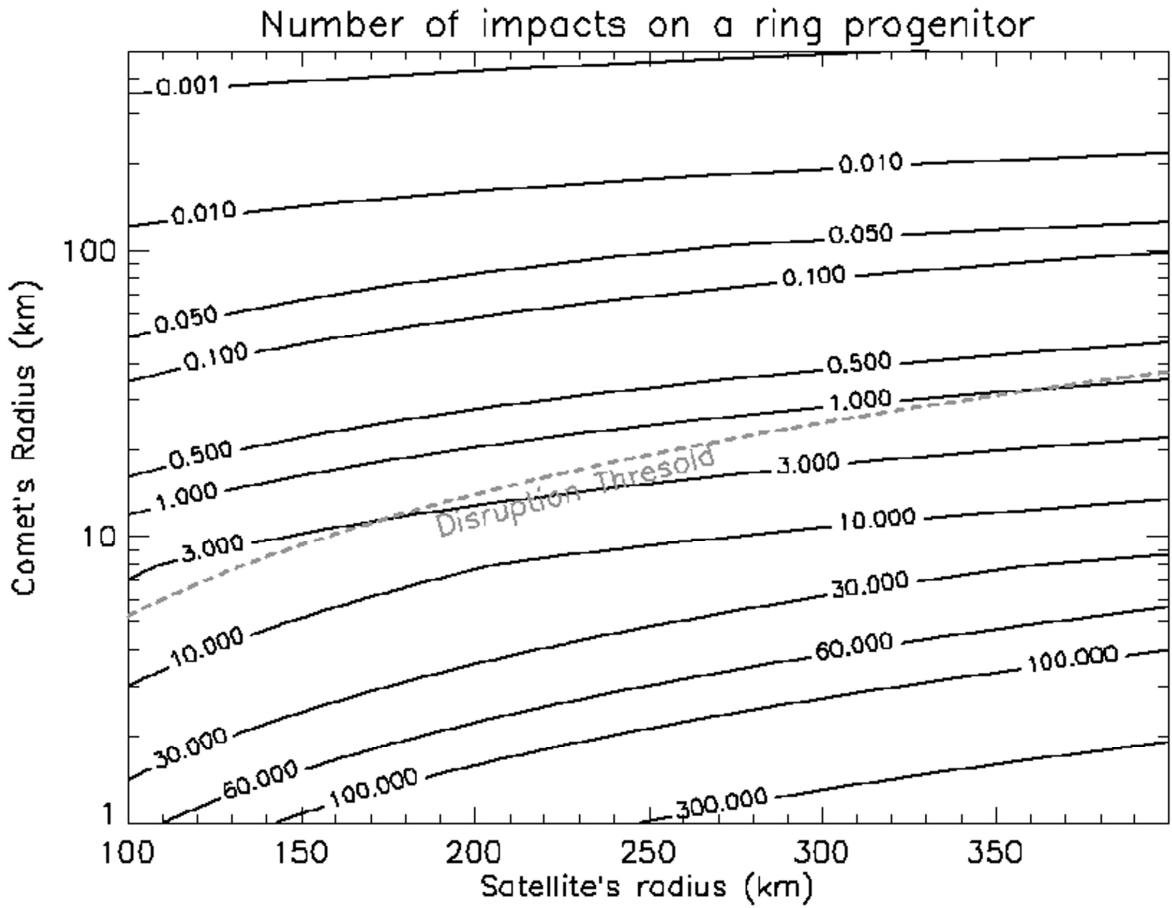

**Figure 8**: Number of comet impacts during the LHB, on a satellite located at 100,000 km from Saturn. The population of impacting comets has the ISD size distribution (see Fig. 2). The dashed grey line shows the minimum comet size for disruption in a single impact, according to the Benz & Asphaug shattering model (1999) for icy bodies. See section **5.2.3** for details.

Given that all tools are set, it is also interesting to compute the "number of destructions" that Saturn's current satellites would have suffered. They are reported in Table 3. For comparison with previously published works, we also report the results obtained adopting two additional criteria for satellite disruption: those of Melosh (1989) and Zahnle et al., (2003). In these papers it was assumed that a satellite is destroyed if a crater is formed with a diameter equal to the satellite's diameter, and the Melosh (1989) and Zahnle (2003) models are, in fact, two different scaling laws to convert from projectile-size to crater-size. More significant than the number of destructive impacts is the probability $P_s$ that a satellite avoids all disruptive collisions. This is computed as follows: Time is divided into N equal small time steps. For each time step i the probability $N_i$ of a disruptive impact is computed as before. Thus, the probability for not having a disruptive impact during the time step is 1-$N_i$ (the time step is chosen short enough so that $N_i$ <1 for any i). So the total probability for not having a disruptive impact during the full evolution of the system is :

$$P_s = \prod_{i=1}^{N}(1-N_i) \quad \textbf{Eq.11}$$

Now, assuming a uniform probability distribution, any time step could be arbitrarily divided into η equally spaced intervals, so that $(1-N_i)$ could be replaced by $(1-N_i/\eta)^\eta$ with η being an arbitrarily large number. In the limit where η tends to infinity, this result tends to $e^{-N_i}$. So finally, the final survival probability is simply given by :

$$P_s = \prod_{i=1}^{n} e^{-N_i} = e^{-N_{total}} \quad \textbf{Eq.12}$$

where $N_{total}$ is the total number of impacts on a satellite, as computed before. These probabilities are also reported in Table 3 (conditions of destruction are displayed in Table 3 's legend). Using any of the three criteria for satellite disruption, we find that Iapetus, Titan, Phoebe, Dione, Rhea, and Tethys would survive with a high probability (note however that Phoebe may be a captured satellite and so should not have formed in Saturn's sub-nebula, see Johnson & Lunine 2005 and Porco et al., 2005) . The case of Enceladus is less clear because, depending on the disruption criterion adopted, it could have a survival chance between 37% (Zahnle model) and 71% (Benz & Asphaug model). Conversely, Mimas and all the satellites interior to its orbit have a very small probability of having escaped a disruptive collision. Most likely all these satellites reaccreted after the LHB from the debris of progenitor satellites.

These results compare well with the Zahnle et al., (2003) results for the case with the impactor distribution "A" and scaled to Iapetus' surface (sixth column in Table 5 of Zahnle et al. 2003), which also predicts that Mimas would be the only big satellite of Saturn destroyed since the formation of Iapetus' surface, whereas Enceladus is a more ambiguous case. However, we note that Smith et al. (1982) predict a much larger destruction rate, in which Enceladus, and even Tethys, are unlikely to survive. We note here that whatever the model and the assumptions (present work, Zahnle et al., 2003, Smith et al., 1982), Mimas is expected to have been destroyed at least once during Solar System history. Only in the Lissauer et al. (1988) work is Mimas expected to have survived, but it seems that the Lissauer et al. (1988) study suffers a number small problems, leading to low disruptions frequencies, in particular by assuming a much too large value of $V_\infty$ (see comments by Zahnle et al., 2003).

**[ PUT TABLE 3 HERE ]**

# 6. Discussion

We have shown that the enormous flux of small bodies during the Late Heavy Bombardment could have been responsible for the origin of Saturn's rings, either via tidal splitting of large comets, or via the collisional destruction of a primitive satellite in the Roche Zone. We now try to compare the pros and cons of these two scenarios with respect to two extreme characteristics: (i) the low abundance of silicates in Saturn's rings and (ii) the obvious fact that only Saturn has a massive ring system.

## 6.1 The problem of silicates

Saturn's rings are mainly composed of pure water ice, with very few silicates (Cuzzi & Estrada 1998, Poulet et al. 2003, Nicholson et al. 2005). However, some silicates may be hidden in the bulk of the massive B ring (Esposito et al., 2007, Stewart et al. 2007). Are there means, in the frame of the two scenarios discussed above, to avoid the presence of a large fraction of silicates in the ring? At first sight, we would expect that both scenarios would produce rings with a much larger fraction of silicates than inferred from observations, because both satellites and comets typically have a larger silicate/ice ratio of order unity (Johnson & Lunine 2005).

However, we have seen (section 5) that big planetesimals (comets), or big satellites, are required to be progenitors of the rings in both scenarios. These big bodies could be differentiated: for example, Enceladus, a body with a radius of only 250 km, seems to be differentiated (Schubert et al. 2007, Thomas et al., 2007). This is the case for Dione (500 km in radius; Thomas et al. 2007) as well. However, generally speaking, it is not known whether outer Solar System planetesimals with radii of hundreds of kilometres are differentiated, although some models suggest it is indeed the case for the biggest ones (McKinnon et al. 2008). In a differentiated body, ice and rock are segregated from each other: denser material is concentrated in the body's core, whereas the mantle is made of lower density material, like ice. Therefore a differentiated progenitor might be the solution for forming a ring system of pure ice, provided that a process exists to separate the ice in the mantle from the silicates embedded in the core, and get rid of the latter. We investigate this question for the two scenarios.

### 6.1.1 The tidal splitting scenario

In the model of big comets disrupted by tides, a separation between the core and the mantle material could be a natural outcome of the tidal splitting process. Indeed, after the splitting of the body, the material most easily trapped on bound orbits is that which is the furthest away from the body's center of mass on the planet-facing side. Indeed, whereas the center of mass has a positive total energy, and therefore travels on a hyperbolic orbit, the surface material has a small energy difference, proportional to the distance from the center of the body (see Eq. 6), that may result in bound orbits for the surface fragments. So we expect that the tidal splitting scenario leads to the preferential capture of surface and mantle material, with a high ice/silicate ratio, whereas the silicate core is lost to unbound orbit. Of course all this would need to be quantified with a numerical model of tidal splitting of a differentiated body.

### 6.1.2 The destroyed satellite scenario

For the case of the satellite destruction scenario, some interesting solutions may also exist to explain the silicate deficit. Whereas a satellite with a few Mimas masses may be hit a few times by 25-km comets and be completely destroyed, Fig. 8 shows that it may be hit several hundred times by 1-km comets (or smaller). A satellite larger than 200 km could be differentiated and be composed of a dense silicate core surrounded by an icy mantle. If the satellite is differentiated, the intense cometary flux may have "peeled off" the satellite of its icy shell, and left a disk of debris surrounding a "naked core" containing the initial silicate content of the satellite. It may be possible that in such an event, the core survives undisrupted, whereas the mantle is shattered away, because the impact shock wave is reflected at the boundary between the core and mantle. This is indeed what happens in simulations of formation of the Moon or of the Pluto-Charon system (Canup 2005), in which the target's core appears to remain undisrupted after the impact, whereas part of its mantle is scattered into space. Other studies (Asphaug et al., 2006) show that in planetary embryo collisions, the outer layers are preferentially stripped off. So, as far as we know, it is not impossible that the icy debris would be put in orbit around Saturn, forming a narrow and dense ring, in the middle of which the "naked" core would be embedded. So, after the destruction, we could expect to have the satellite's core surrounded by two rings (inner and outer) on both sides of the core's orbit. The total mass of the debris rings (inner+outer) would be comparable to the core's mass. Such a configuration could be very unstable, because the core would gravitationally interact with the icy ring, and might remove it from the Roche Zone. However, this is a complex scenario that deserves detailed modeling in order to infer whether the naked core would migrate away from the ring region due to the tidal interaction with the debris. This should be considered as a quite hypothetical way to eliminate silicates that will be investigated in the future. So for the moment there is no obvious way of eliminating silicates in the frame of this scenario.

## 6.2 Massive rings around other giant planets?

One of the most puzzling mysteries about giant planets is why Saturn is the only planet with a massive ring-system A scenario of Saturn ring formation should give an answer to this question, and this issue can be used as a criterion for its validation.

### 6.2.1 The tidal splitting scenario

The scenario of tidal splitting suffers from a severe problem : we have shown in section 5.1 that the capture rate around Saturn is the *lowest* among giant planets, because of the combined effect of Saturn's low density, its large distance from the original planetesimal disk compared to Uranus or Neptune, and also its low mass compared to Jupiter. Whatever the size distribution of incoming objects, Saturn is expected to receive the lowest quantity of material. So we are in quite a paradoxical situation with the tidal splitting scenario: if Saturn acquired its massive ring system through this mechanism, we would expect that the other three giant planets would have rings even more massive than Saturn! How can we solve this problem?

We have already discussed in Sect. 4.1 possible reasons for which the amount of material captured from splitting events in the ring region could be much smaller than we calculated. More specific studies are required to clarify this issue. But we do not see a priori any reason for which the capture efficiency would be high at Saturn and low at Jupiter, Uranus and Neptune. So, either the capture efficiency is high for all planets and all planets have massive rings (which is obviously not the case) or the capture efficiency is low for all planets, and Saturn's rings could not form by the tidal disruption mechanism. We note, however, that if Jupiter had icy rings, they probably would not survive over the age of the Solar System (Watson et al. 1962), so we cannot discard the idea that Jupiter could have had massive rings in the past.

### 6.2.2 The destroyed satellite scenario

For this model, a key aspect is the survival of the satellite in the planet's Roche Zone for 700 My, up to the time of the LHB. As explained in section 5.2.2, tidal interaction with the host planet may lead to a rapid migration toward the planet (if the satellite starts below the synchronous orbit) or away from the planet (if the satellite starts above the synchronous orbit). However, the two configurations (above and below the synchronous orbit) are not equivalent : due to the term proportional to $a^{-11/2}$ in the equation for the migration rate (Eq. 9), the orbital decay of a satellite below the synchronous orbit is an accelerating mechanism, leading to a rapid fall onto the planet. So a critical parameter to estimate if a satellite can survive 700 My in the planet's Roche Zone is the position of the planet's Synchronous Orbit relative to the Roche Limit. The synchronous orbits of Jupiter, Saturn, Uranus and Neptune are located respectively at $R_{synch}$=2.24, 1.86, 3.22, 3.36 planet's radii. Since the classical Roche Zone is located at $R_{Roche}$=2.456$\times R_p \times (\rho_{planet}/\rho_{material})^{1/3}$ (where $R_p$ is the planetary radius, $\rho_{planet}$ is the planet's density and $\rho_{satellite}$ is the satellite's density Roche 1847, Chandrasekhar 1969), and assuming satellites with material density of 1g/cm$^3$, we obtain $R_{Roche}$=2.70, 2.24, 2.79, 2.89 planetary radii for Jupiter, Saturn, Uranus and Neptune respectively. We observe that only Jupiter and Saturn have their Synchronous Orbit below their Roche Limit whereas Uranus and Neptune have their synchronous orbit above their Roche Limit. Therefore it is almost impossible for Uranus and Neptune to maintain a Mimas-sized satellite inside their Roche Zone up to the onset of the LHB, as illustrated in Fig. 9. One may note that today Neptune shelters a number of moderate-size satellites inside its Roche Zone (Naiad with D ~ 58 km, Thalassa D ~80 km), Despina D~ 148 km, Galatea D ~158 km, and Larissa D

~190 km, where D is the mean diameter), that would have been disrupted during the LHB (the expected numbers of destructive impacts are 8.0, 7.4, 6.6, 5.2, and 3.6 for Naiad, Thalassa, Despina, Galatea and Larissa, respectively, assuming their present locations) and would not have reaccreted if they were at their present locations at the time of the LHB. A recent study on the orbital evolution of Neptune's satellites (Zhang & Hamilton 2008) shows that due to their small masses, they tidally migrate on long timescales (~Gy), and it is plausible that at the time of LHB these satellites would have been above the Roche Limit and below Neptune's synchronous orbit (as seen in Fig. 2 of Zhang & Hamilton 2008). A moon with a mass comparable, or larger than, Mimas ($\Leftrightarrow$ D ~ 400 km) would fall onto Neptune in a few $10^8$ years (e.g. Fig. 9). In consequence average-sized moons, like those today around Neptune, could have formed below the Synchronous orbit (and above the Roche Limit), then been disrupted while they were still beyond the Roche Limit, reaccreted, and finally tidally migrated inward to their present locations within Neptune's Roche Zone. Such a scenario is supported by the results of Zhang & Hamilton (2008). A similar scenario may hold also for Uranus' moons Ophelia (D=32 km) and Cordelia (D=26 km).

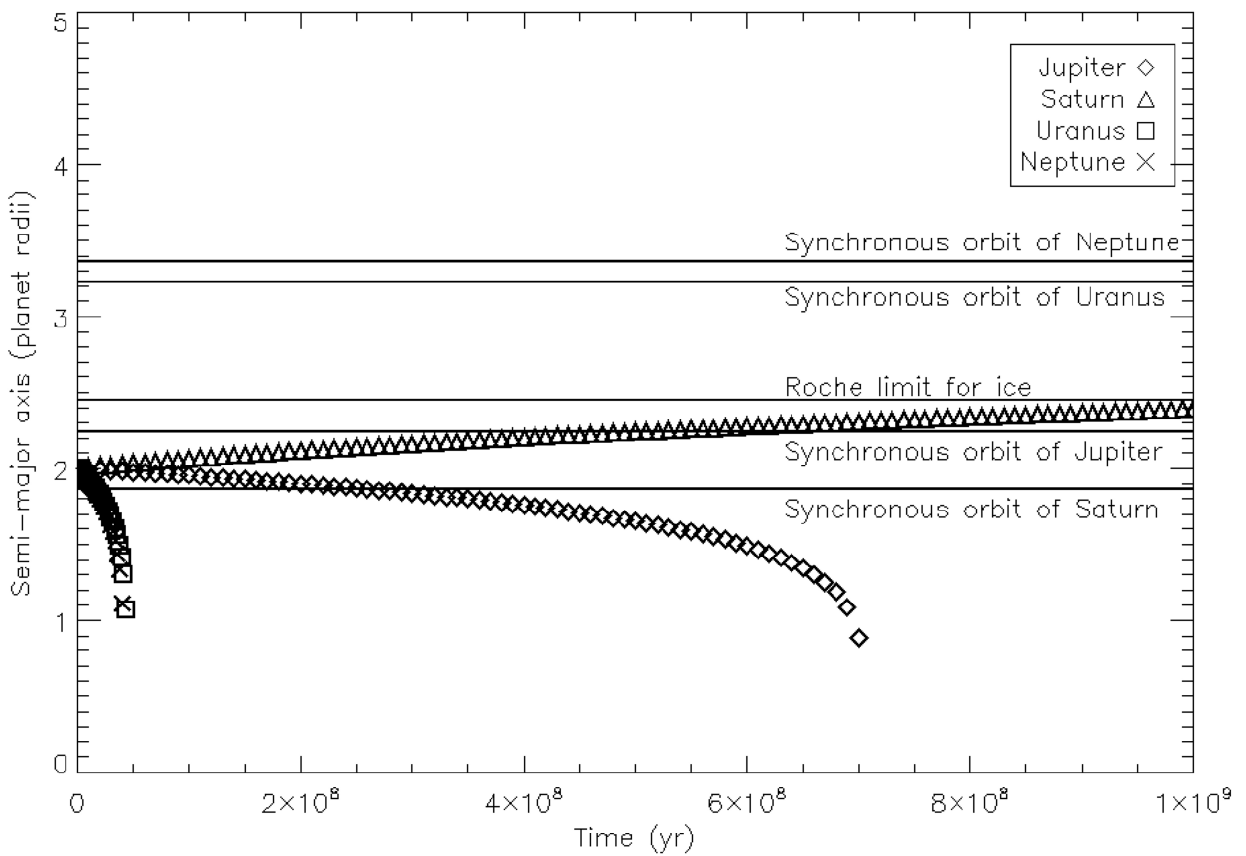

**Figure 9**: Tidal migration of a 3 Mimas' mass satellite, starting at 2 planetary radii for all four giant planets. $Q=10^5$ has been assumed for all planets.

The case of Jupiter is less clear. Jupiter's synchronous orbit is slightly below its Roche Limit for ice, and it seems, a priori, to be in a similar situation as Saturn. However, $R_{synch}/R_{Roche}$ is 0.91 for Jupiter whereas it is 0.76 for Saturn. (assuming that the satellites have the density of water ice) Consequently, the region in the planet's Roche Zone which is above the Synchronous Orbit is somewhat wider for Saturn. Finally, we note that the mean density of Jupiter's satellites is higher than for Saturn's satellites and is more representative of silicates than of ice. By setting the satellite density to 2g/cm$^3$ we obtain a Roche radius for silicates at Jupiter of ~2.14 $R_p$, which is below Jupiter's synchronous orbit, at 2.24 $R_p$. However, the large mass of Jupiter makes the migration slower and whereas the four main Galilean satellites have a high density, Amathea's density is only 0.857±0.099 g/cm$^3$ (Anderson et al., 2005), somewhat softening the above arguments. So no definitive conclusion can be drawn for Jupiter.

In conclusion, it seems that in the frame of the satellite disruption scenario, Saturn is the planet with the most obvious chance to shelter a massive satellite (a couple of Mimas' masses) inside the Roche region for a long time, whereas this is almost impossible for Uranus and Neptune. The case of Jupiter is somewhat intermediate between Saturn's case and Uranus-Neptune's case.

# 7. Conclusion: the LHB is the "sweet moment"

We have studied the conditions to form Saturn's massive ring system during the Late Heavy Bombardment (LHB). In particular, we have quantified the probabilities associated with two often invoked scenarios: (i) the tidal splitting of a massive comet and (ii) the destruction of a satellite located in the Roche Zone. The impact rates on a satellite and the probability of passing comets within the planet's Roche radius have been evaluated from numerical simulations of the LHB, in the frame of the *Nice Model* (Tsiganis et al., 2005, Gomes et al., 2005, Morbidelli et al., 2005). The size distribution of the planetesimals involved in the LHB has been estimated from the crater record on Iapetus, and is consistent with size distributions derived from studies of the evolution of the Kuiper Belt, Scattered Disk and Oort Cloud from the primitive disk (Charnoz & Morbidelli 2007). We have shown that the flux of trans-Neptunian planetesimals through Saturn's system during the LHB was high enough for both scenarios to be valid.

For the two scenarios we have given a list of conditions and caveats. It appears that :
- For the tidal splitting scenario: the major caveat is that Saturn appears to be the planet that receives the least amount of material. Thus, in the frame of the LHB, if Saturn acquired its massive rings through delivery of material by tidal splitting of passing comets, we would expect the other three giant planets to also have massive rings, which is obviously not the case. This conundrum is independent of models of tidal capture, and comes simply from the specific dynamics of planetesimals from the trans-Neptunian disk (it is even independent of the details of the dynamics of the planets in the Nice model). In contrast, this model offers a natural explanation for the missing silicates problem, since the capture of material is dominated by the tidal splitting of massive objects, which would presumably be differentiated. In fact, a differentiated body would have its outer mantle tidally stripped more easily than its high density core.
- For the model of a destroyed satellite, we have shown that according to our current knowledge of satellite formation (see the work of Canup & Ward 2006), it could be possible to bring a satellite inside the planet's Roche Zone through type-I migration during the early phases of giant planet formation. If the Synchronous orbit is below the Roche Limit, then it is possible to maintain the satellite inside the Roche Limit for 700 My (the time of the LHB onset). Saturn is the planet where such an event is the most probable due to its synchronous orbit being well below its Roche Limit. Then, at the time of the LHB, a ~20-30 km comet could destroy the satellite with a probability larger than 75% (depending on the satellite's size, see Fig. 8). On the flip side, there is no obvious way to eliminate the silicates ejected from the disrupted satellite, so that the missing silicate problem remains open, although we suggest a speculative mechanism (see section **6.2.2**)
- The LHB also has interesting implications for Saturn's satellites: Mimas and all smaller satellites would not have survived the LHB. These satellites could be in fact bodies that re-accreted after the LHB. Enceladus has a roughly 50% chance of survival. Iapetus would have survived the LHB, and its big basins may have formed during this period.

In conclusion, we think that the LHB is the `sweet moment' for formation of a massive ring system around Saturn. On the basis of simple arguments we find that the scenario of a destroyed satellite gives a substantially more coherent explanation of the fact that Saturn' rings are unique by mass in the Solar System. Nevertheless, both scenarios deserve and require more detailed numerical modelling of tidal and collisional disruption in order to be able to give a more definite answer for the origin of Saturn's rings.

**Acknowledgements:**
The authors would like to thank M.J.Burchell and A. Lightwing for very enlightening comments on the physics of ice fracture, as well as K. Holsapple and P.Michel for useful discussions about the physics of tidal disruption. We would like also to thank our reviewers J. Colwell and A. Harris whose comments increased the quality of the paper. This work was supported by Université Paris 7 Denis Diderot and the NASA Planetary Geology and Geophysics Program.

# TABLES

| Implanted Mass | JUPITER | SATURN | URANUS | NEPTUNE |
|---|---|---|---|---|
| Below the Planet Hill Radius | Mean mass = 245.591<br>Median mass = 139.978<br>1 sigma mass dispersion = 212.126<br><br>Mean Angular Momentum = 0.541337<br><br>Median Momentum = -0.0628934<br><br>Angular Momentum standard deviation = 23.1733 | Mean mass = 47.9874<br>Median mass = 15.5280<br>1 sigma mass dispersion = 85.0293<br><br>Mean Angular Momentum = -1.82926<br><br>Median Momentum = -0.0270548<br><br>Angular Momentum standard deviation = 7.56165 | Mean mass = 132.369<br>Median mass = 87.5163<br>1 sigma mass dispersion = 169.888<br><br>Mean Momentum = -1.98800<br><br>Median Momentum = 0.113855<br><br>Angular Momentum standard deviation = 35.9674 | Mean mass = 247.969<br>Median mass = 207.673<br>1 sigma mass dispersion = 216.153<br><br>Mean Momentum = -5.32988<br><br>Median Momentum = 0.181543<br><br>Angular Momentum standard deviation = 46.3779 |
| With apocenter below 50 planet radii | Mean mass = 0.<br>Median mass = 0.<br>*0/100 events* | Mean mass = 6.50<br>Median mass = 0.<br>*3/100 events* | Mean mass = 3.76<br>Median mass = 0.<br>*15/100 events* | Mean mass = 29.02<br>Median mass = 0.<br>*35/100 events* |
| With apocenter below 20 planet radii | Mean mass = 0.<br>Median mass = 0.<br>*0/100 events* | Mean mass = 0.<br>Median mass = 0.<br>*0/100 events* | Mean mass = 2.57<br>Median mass = 0.0<br>*4/100 events* | Mean mass = 12.25<br>Median mass = 0.0<br>*4/100 events* |

**TABLE 2:** Statistics on masses and angular momentum implanted below $R_{stab}$ for 100 simulations of comets passing around each of the giant planets. Masses are in units of Mimas' mass, distance in units of the planet's radius, and GM=1 for each planet.

| Name | Distance (km) | Radius (km) | Number of Disruptions : B&A(♣) | Survival probability B&A model | Number of Disruptions : Melosh crater model (♦) | Survival Probability Melosh Model | Number of Disruptions : Zahnle crater model(♠) | Survival Probability Zahnle Model |
|---|---|---|---|---|---|---|---|---|
| Ring progenitor | 100000 | 320 | 1.02 | 0.36 | 4.07 | 0.017 | 7.09 | 0.00083 |
| Pan | 133583 | 14 | 3.13 | 0.044 | 37.37 | 0. | 25.15 | 0. |
| Atlas | 137700 | 15 | 2.97 | 0.051 | 32.25 | 0. | 22.19 | 0. |
| Prometheus | 139400 | 43 | 2.62 | 0.073 | 20.86 | 0. | 18.29 | 0. |
| Pandora | 141700 | 40 | 2.58 | 0.076 | 19.98 | 0. | 17.26 | 0. |
| Epimetheus | 151400 | 56 | 2.28 | 0.10 | 11.42 | 0. | 10.72 | 0. |
| Janus | 151500 | 89 | 2.19 | 0.11 | 10.63 | 0. | 11.08 | 0. |
| Mimas | 185600 | 198 | 0.73 | 0.48 | 3.04 | 0.05 | 4.46 | 0.011 |
| Enceladus | 238100 | 252 | 0.34 | 0.71 | 0.62 | 0.54 | 0.992 | 0.37 |
| Telesto | 294700 | 12 | 1.08 | 0.33 | 4.31 | 0.013 | 2.86 | 0.057 |
| Calypso | 294700 | 10 | 1.10 | 0.33 | 3.92 | 0.020 | 2.52 | 0.08 |
| Tethys | 294700 | 533 | 0.08 | 0.92 | 0.28 | 0.75 | 0.59 | 0.55 |
| Dione | 377400 | 561 | 0.05 | 0.95 | 0.098 | 0.91 | 0.21 | 0.81 |
| Rhea | 527100 | 764 | 0.01 | 0.99 | 0.039 | 0.96 | 0.10 | 0.90 |
| Hyperion | 1464099 | 146 | 0.06 | 0.94 | 0.74 | 0.47 | 0.87 | 0.41 |
| Titan | 1221850 | 2575 | 0.00013 | 0.9998 | 0.00011 | 0.999887 | 0.00062 | 0.9994 |
| Iapetus | 3560800 | 736 | 0.00112 | 0.9988 | 0.0061 | 0.993924 | 0.019 | 0.9806 |
| Phoebe | 12944300 | 110 | 0.019 | 0.981 | 0.07 | 0.932 | 0.085 | 0.9184 |

**Table 3**: Number of destructions of Saturn's satellites during the LHB, using different disruption models with the Iapetus-scaled size distribution of the primordial Kuiper-Belt population (see figure 2). (♣) The specific destruction energy is taken from Benz & Asphaug (1999); destruction is defined so that the mass of the largest fragment is less than 50% of the parent's body mass (♦) Crater with same diameter as the parent body, using the crater scaling law of Melosh (1989) (♠) Crater with same diameter as the parent body, using the crater scaling law of Zahnle et al. (2003). Survival probabilities are computed according to Eq. 12. Note that the survival probability is simply exp(-Number of Disruptions), see section 5.2.3.